\documentclass[12pt]{article}
\usepackage{epsfig,amssymb,amsmath,psfrag,cite,hyperref}

\usepackage{color}

\newcommand{\cN}{{\cal N}}

\newcommand{\cO}{{\cal O}}

\def \be  {\begin{equation}}
\def \ee  {\end{equation}}
\def \ba  {\begin{eqnarray}}
\def \ea  {\end{eqnarray}}

\def \tr {\mathop{\rm tr}\nolimits}

\def \cO{\mathcal{O}}

\begin{document}

\thispagestyle{empty}

\null\vskip-43pt \hfill
\begin{minipage}[t]{30mm}
	DCPT-17/13
\end{minipage}

\null\vskip-12pt \hfill  \\
\null\vskip-12pt \hfill   \\

\vskip2.2truecm
\begin{center}
\vskip 0.2truecm {\Large\bf
{\Large Quantum Gravity from Conformal Field Theory}
}\\
\vskip 1truecm
{\bf F.~Aprile${}^{1,2}$, J.~M. Drummond${}^{2}$, P.~Heslop${}^{3}$, H.~Paul${}^{2}$ \\
}

\vskip 0.4truecm
 
{\it
${}^{1}$ Mathematical Sciences and STAG Research Centre, \\
University of Southampton, Highfield, SO17 1BJ,\\
\vskip .2truecm }
\vskip .2truecm
{\it
${}^{2}$ School of Physics and Astronomy and STAG Research Centre, \\
University of Southampton,
 Highfield,  SO17 1BJ,\\
\vskip .2truecm                        }
\vskip .2truecm
{\it
${}^{3}$ Mathematics Department, Durham University, \\
Science Laboratories, South Rd, Durham DH1 3LE \vskip .2truecm                        }
\end{center}

\vskip 1truecm 
\centerline{\bf Abstract} 

We bootstrap loop corrections to AdS${}_5$ supergravity amplitudes by enforcing the consistency of the known classical results with the operator product expansion of $\mathcal{N}=4$ super Yang-Mills theory. In particular this yields much new information on the spectrum of double-trace operators which can then be used, in combination with superconformal symmetry and crossing symmetry, to obtain a prediction for the one-loop amplitude for four graviton multiplets in AdS. This in turn yields further new results on subleading $O(1/N^4)$ corrections to certain double-trace anomalous dimensions.

\medskip

 \noindent

\newpage
\setcounter{page}{1}\setcounter{footnote}{0}


\section{Introduction and summary}

The best understood example of the AdS/CFT correspondence~\cite{hep-th/9711200,hep-th/9802109,hep-th/9802150} equates IIB string theory on AdS${}_5 \times$S${}^5$ to $\cal N$=4, $SU(N)$ super Yang-Mills theory. 
Gauge invariant operators in $\cN=4$ SYM  are related to string states in IIB, 
correlation functions of gauge invariant operators are related to AdS amplitudes and the free dimensionless parameters on both sides are related as $g_{YM}^2 = g_S$,  $ (g_{YM}^2 N)^{-1/4}= l_S/L$ where $g_{YM}$ is the Yang-Mills coupling constant, and $l_S/L$ is the ratio of the string length to the AdS radius.

The simplest states to consider on the string theory side are those belonging to the AdS${}_5$ graviton supermultiplet and its Kaluza-Klein partners. These states correspond to operators in half-BPS multiplets in the gauge theory. We denote the superconformal primaries of the half-BPS multiplets by $\cO_{p}$, the gauge invariant product of $p$ copies of one of the  scalars together operators in the same  representation of the internal symmetry group $SU(4)$. The case $p=2$ corresponds to the graviton multiplet itself.

The two-point and three-point amplitudes of half-BPS states are independent of the coupling~\cite{hep-th/9811047}, and thus the same at weak coupling and strong coupling, one of the early tests of AdS/CFT~\cite{hep-th/9806074}. Thus the first non-trivial supergravity amplitudes of the half-BPS states appear at four points. The simplest ones are those involving states from the graviton supermultiplet itself and they are related to the four-point correlators of operators in the stress-tensor supermultiplet. We denote the correlator of the superconformal primaries by $\langle 2222 \rangle = \langle \cO_{2}\cO_{2}\cO_{2}\cO_{2} \rangle$. This four-point function has  immense interest, not least as one can extract from it information about non-protected operators, and it has been the object of a huge amount of research throughout the intervening time, both perturbatively~\cite{GonzalezRey:1998tk,Eden:1998hh,Eden:1999kh,Eden:2000mv,Bianchi:2000hn,Eden:2011we,Eden:2012tu,Ambrosio:2013pba,Drummond:2013nda,Bourjaily:2015bpz,1609.00007} and in the supergravity approximation~\cite{Liu:1998ty, hep-th/9903196,hep-th/9911222,hep-th/0002170,hep-th/0212116}.

At large $N$, keeping the `tHooft coupling $\lambda=g_{YM}^2 N$ fixed, the four-point correlator $\langle 2222 \rangle$ has an expansion of the form
$$\langle 2222 \rangle = \langle 2222 \rangle^{(0)} +a \langle 2222 \rangle^{(1)}[\lambda]+a^2 \langle 2222 \rangle^{(2)}[\lambda] + \dots 
$$
where $a=1/(N^2-1)$.
This corresponds to a loop expansion on the string theory side so that  $\langle 2222 \rangle^{(0)}$ is dual to the free (disconnected) string amplitude and is independent of $l_S$. The next term, $\langle 2222 \rangle^{(1)}[\lambda]$, is dual to the tree-level amplitude, and $\langle 2222 \rangle^{(2)}[\lambda]$ is dual to the one-loop string amplitude, both of which depend on $l_S$ or equivalently $\lambda$. 
In a perturbative expansion for small $\lambda$ the tree-level amplitude $\langle 2222 \rangle^{(1)}[\lambda]$ is now known to ten loops in terms of conformal integrals \cite{1609.00007} and to three loops in terms of explicit polylogarithms \cite{Drummond:2013nda}. Here however we are interested in the expansion at large $\lambda$ corresponding to small $l_S/L$. The leading term in this large $\lambda$ limit then corresponds to a tree-level supergravity amplitude. This computation was performed on the supergravity side  in AdS/CFT~\cite{hep-th/9903196,hep-th/0002170}. More recently the first few $l_S/L$ corrections to this result  were computed explicitly in Mellin space~\cite{1411.1675}. They first correction is of order $\lambda^{-3/2}$, corresponding to order $l_s^6$ or order $\alpha'^3$. 

Until now however there has been no study of string loop corrections $\langle 2222 \rangle^{(2)}$ (although see~\cite{1612.03891} where a study of loop corrections in  more general AdS context was initiated). In this paper we take this step. We give a precise prediction for the leading term in $1/\lambda$ to $\langle 2222 \rangle^{(2)}[\lambda]$ dual to the first loop correction to the four-graviton superamplitude. We will denote the leading terms in the large $\lambda$ expansion simply by $\langle 2222 \rangle^{(n)}$ from now on.  

We perform our analysis by analysing the OPE decomposition of the tree-level supergravity result $\langle 2222 \rangle^{(1)}$ together with the recently found supergravity results for arbitrary charge correlators of the form  $\langle ppqq \rangle^{(1)}$~\cite{1608.06624} (see also previous work by~\cite{hep-th/0212116,hep-th/0301058,hep-th/0601148,0709.1365,0811.2320,1106.0630}). With the information on the spectrum and three-point functions thus obtained we are able to employ an analytic bootstrap of the type recently employed in weak coupling studies of both correlators and scattering amplitudes in $\mathcal{N}=4$ SYM \cite{Dixon:2011pw,Drummond:2013nda,Dixon:2014voa,Drummond:2014ffa,Dixon:2016nkn}.

Consider the expansion of  the four-point function in superconformal blocks and its relation to the OPE~\cite{hep-th/0112251}. In the limit $x_{12}^2\rightarrow 0$, $\langle 2222 \rangle^{(2)}$ has a leading divergence  $\log^2 x_{12}^2$  whose coefficient function depends only on data of lower order in the $1/N$ expansion. 
Specifically, it is completely determined by the following data
$$\frac{1}{2}  \sum_{i=1}^{t-1} (C_{22}^{tli})^2 (\eta_{tli}^{(1)})^2\ 
$$
for $t\geq 2, l\geq 0$. Here $C_{22}^{tli}$ are zeroth order three-point functions of two stress-tensor multiplets and a double trace $SU(4)$ singlet operator of twist $2t$, spin $l$, and $\eta_{tli}^{(1)}$ is (half) the operator's  $O(1/N^2)$ anomalous dimension.
There are precisely $t-1$ double trace operators, $K_{tli}$, for each $t,l$, with  $i=1,2,..,t{-}1$ labelling the different operators. They  are linear combinations of superconformal primary operators of the schematic form $\cO_{p} \partial^l \Box^{t-p} \cO_{p}$ for $p= 2,3,..t$. We have assumed here that all unprotected single trace operators have disappeared from the spectrum in this limit (they correspond to string states with very large masses) and in addition that triple trace operators are suppressed by a further order of $1/N^2$.

So the $\log^2 x_{12}^2$ coefficient is determined in terms of this lower order data, but unfortunately this data can not be extracted directly from the lower order charge two correlators alone due to mixing: there are $t-1$ operators with the same quantum numbers. 
However it turns out that we {\em can} extract this data from  the correlators $\langle ppqq \rangle$ at leading and subleading order in $1/N^2$  for arbitrary $p,q$. The $t(t-1)/2$ correlators with $p,q =2.. t$ contain data involving $K_{tli}$, specifically
\begin{align*}
\sum_{i=1}^{t-1} C_{pp}^{i}C_{qq}^{i} \eta_{i}^{(1)}, \qquad
\sum_{i=1}^{t-1} C_{pp}^{i}C_{qq}^{i} , \qquad p,q=2,..,t
\end{align*}
where we have suppressed the dependence on twist and spin. This gives $t(t-1)$ equations, precisely equal to the number of unknowns:  the $(t-1)^2$ 3-point functions  $C_{pp}^{i}$, and the $t-1$ anomalous dimensions. 
We can thus solve these equations to obtain the 3-point functions and anomalous dimensions. The full analytic formulae for the $O(1/N^2)$ anomalous dimensions, for all $t,l,i$  we obtain via this procedure is given by the compact formula
\be
\eta_{t,l,i} = -\frac{2(t-1)_4 (l+t)_{4}}{(l+2i-1)_{6}}\,,\label{eta}
\ee
and the corresponding formulae for $C_{22}^{tli} $ is displayed below  in~(\ref{3pnt}).

With these results we can then completely determine the full $\log^2 x_{12}^2 $ coefficient of the string loop amplitude $\langle 2222 \rangle^{(2)}$ (see~\eqref{dbldisc}). 
Then having obtained the full $\log^2 x_{12}^2 $ coefficient we complete it to a full crossing-symmetric function. The completion is unique up to crossing symmetric functions with no $\log^2 x_{12}^2$ singularity which our technique can never capture. One such correction is of the form $\alpha \bar D_{4444}$ for an unfixed $\alpha$. Other possible correction terms can be written as $\bar D$ functions with higher values of the parameters (see~\cite{1410.4717}). Our general prediction for  $\langle 2222 \rangle^{(2)}$ is given in~\eqref{forientations} and the following equations.

Finally, having obtained the full correlator we can now in turn extract new data from it. In particular, we can extract the gravity loop corrected anomalous dimensions of twist 4 operators (for higher twist operators we expect mixing with triple trace operators to spoil this). For $l\geq 2$, we find
$$ \eta^{(2)}_{2,l} = \frac{1344 (l-7) (l+14)}{(l-1) (l+1)^2 (l+6)^2 (l+8)} -\frac{2304 (2 l+7)}{(l+1)^3 (l+6)^3}\,.$$

The paper proceeds as follows.
In section~\ref{sec2} we introduce the correlators in more detail, both the free theory and the general structure arising from superconformal symmetry.
In section~\ref{sec3} we introduce the OPE and super conformal partial waves. In section~\ref{sec4} we discuss the supergravity limit and the operators we expect to remain in the spectrum: we display our results for the 3-point functions and anomalous dimensions extracted from the supergravity data. Section~\ref{sec5} contains details o the resummation of this data to obtain the $\log^2 x_{12}^2$ coefficient. In section~\ref{sec6} we complete this to the full correlator, then in section~\ref{sec7} we extract the anomalous dimensions at $O(1/N^4)$ from this and in the conclusions we discuss our results. The details of the superconformal block expansion we give in a short appendix.

\section{Four-point functions of half-BPS multiplets}
\label{sec2}

We would like to investigate the structure of four-point functions of half-BPS multiplets in $\mathcal{N}=4$ super Yang-Mills theory. The superconformal primary operators we consider are single-trace operators constructed from scalars fields,
\be
\mathcal{O}_{p}(x,y) = y^{R_1} \ldots y^{R_p} \tr \bigl(\phi_{R_1} (x) \ldots \phi_{R_p}(x) \bigr)\,,
\label{Oxy}
\ee
where $y^{R}$ denote a set of auxiliary variables transforming in the vector representation of $SO(6)$ and obeying $y^R y^R =0$. All other operators in the supermultiplet can be obtained from (\ref{Oxy}) by applying supersymmetry transformations. 

Since the operators $\mathcal{O}_{p}$ are protected by supersymmetry, their two-point functions and three-point functions are fully described by their free field expressions. In other words their scaling dimensions and OPE coefficients are unrenormalised and therefore independent of the Yang-Mills coupling. The four-point functions of such protected operators are generically coupling dependent however, since unprotected operators can be exchanged in the operator product expansion.

Many results are available in the literature on four-point correlation functions of the operators $\mathcal{O}_{p}$. In perturbation theory explicit results in terms of polylogarithmic functions are available at one and two loops \cite{hep-th/0003096,hep-th/0212116,hep-th/0301058,hep-th/0305060,hep-th/0504061,arXiv:1408.3527,1512.02926}, and (in the planar limit) at three loops \cite{1108.3557,1303.6909,1512.02926} for all possible choices of external charges $p_i$. In terms of conformal integrals, results are available for higher loop orders for the simplest $\langle 2222 \rangle$ case \cite{1201.5329,1512.07912,1609.00007}.

The correlators have also been investigated in the supergravity limit, where they have a dual interpretation as the scattering amplitudes of AdS supergravity fields \cite{hep-th/9903196,hep-th/9911222,hep-th/0002170}. In particular the supermultiplet with primary $\mathcal{O}_{2}$ contains the energy-momentum tensor and is therefore dual to the graviton multiplet in AdS. The higher charge operators correspond to Kaluza-Klein copies coming from the reduction down from ten dimensions on $S^5$. Recently, a beautifully simple formula was proposed which consistently gives the Mellin space representation for the correlation functions with arbitrary charges \cite{1608.06624} in the regime of classical supergravity. Here we will discuss how to exploit the operator product expansion to bootstrap the supergravity loop corrections.

The fact that the operators $\mathcal{O}_{p}$ are half-BPS means that the four-point functions of any operators in the supermultiplets are uniquely determined in terms of the four-point functions of the superconformal primaries,
\be
\langle p_1 p_2 p_3 p_4 \rangle = \langle \mathcal{O}_{p_1}(x_1,y_1) \mathcal{O}_{p_2}(x_2,y_2) \mathcal{O}_{p_3}(x_3,y_3) \mathcal{O}_{p_4}(x_4,y_4) \rangle\,.
\label{4pt}
\ee
The correlation function (\ref{4pt}) is a homogeneous polynomial of degree $p_i$ in the $y_i$ variables. Our primary focus here is the case of four energy-momentum multiplets $\langle{2222}\rangle$. In order to discuss the supergravity loop corrections to this correlator we will also need some results from more general correlators of the form $\langle{ppqq}\rangle$. 

In free field theory the correlation functions can be written as polynomials in the superpropagators
\be
g_{ij} = \frac{y_{ij}^2}{x_{ij}^2}\,,
\ee
where $y_{ij}^2 = y_i \cdot y_j$. It is also useful to introduce conformal and $su(4)$ invariant cross-ratios
\begin{align}
u = x \bar{x}&=\frac{x_{12}^2 x_{34}^2}{x_{13}^2 x_{24}^2} \,, \qquad &v =(1-x)(1-\bar{x})&= \frac{x_{14}^2 x_{23}^2}{x_{13}^2 x_{24}^2}\,, \notag \\
\qquad y \bar y  &= \frac{y_{12}^2 y_{34}^2}{y_{13}^2 y_{24}^2}\,, \qquad& (1-y)(1-\bar y) &= \frac{y_{14}^2 y_{23}^2}{y_{13}^2 y_{24}^2} \,.
\end{align}
and in particular we will interchange freely between $x,\bar x$ and $u,v$ often including both variables in the same formula.
Notice the useful relations betwen these cross ratios and the superpropagators
\begin{align}
	\frac{g_{12}g_{34}}{g_{13}g_{24}}=\frac{y \bar y}{x \bar x}\,, \qquad 	\frac{g_{14}g_{23}}{g_{13}g_{24}}=\frac{(1-y) (1-\bar y)}{(1-x) (1-\bar x)} \,.
\end{align}

The dependence of the four-point functions on the gauge coupling is heavily constrained by superconformal symmetry. To express the constraints imposed by superconformal symmetry it is useful to separate the correlator into a free-field piece and an interacting piece. In the case of the $\langle 2222 \rangle$ correlator we have
\be
\label{G2222genform}
\langle{2222}\rangle = \langle{2222}\rangle_{\rm free} +  \langle{2222}\rangle_{\rm int} \,,
\ee
where the interacting piece is governed by a single function of the two conformal cross-ratios,
\be
\label{2222int}
\langle{2222}\rangle_{\rm int} = g_{13}^2 g_{24}^2 s(x,\bar x;y,\bar y) F(u,v)\,.
\ee
Here we have
\begin{align}
s(x,\bar x;y,\bar y) &= { (x-y)(x-\bar{y})(\bar{x}-y)(\bar{x}-\bar{y})}\,,
\end{align}
which describes the full $y$-dependence of the interacting term. It is the presence of the factor $s(x,\bar x;y,\bar y)$ in the interacting piece which follows from superconformal symmetry and this feature is sometimes referred to as `partial non-renormalisation' \cite{hep-th/0009106}.
Crossing symmetry implies 
\be
F(u,v) = F(v,u) = \frac{1}{u^4}F\biggl(\frac{1}{u},\frac{v}{u}\biggr)\,.
\ee

For later convenience we choose to normalise our correlation function so that the free-field correlator has the form\footnote{In other words we have divided by a factor of $A=4(N^2-1)^2$ compared to more usual conventions as in e.g. \cite{1508.03611}. This amounts to dividing the operator $\mathcal{O}^{(2)}$ by a factor of $\sqrt{2(N^2-1)}$.}
\be
\label{G02222}
\langle{2222}\rangle_{\rm free}= \langle{2222}\rangle^{\rm (0)}_{\rm free} + a \langle2222\rangle^{\rm (1)}_{\rm free}
\ee
with
\begin{align}
 \langle{2222}\rangle^{\rm (0)}_{\rm free} &= \biggl(g_{12}^2g_{34}^2 + g_{13}^2g_{24}^2+ g_{14}^2g_{23}^2\biggr)\,, \notag \\
 \langle2222\rangle^{\rm (1)}_{\rm free} &= 4\biggl(g_{12}g_{34}g_{13}g_{24}+g_{12}g_{34}g_{14}g_{23}+g_{13}g_{24}g_{14}g_{23} \biggr)
\end{align}
and
\be
\label{G2222ABcoeffs}
a = \frac{1}{N^2-1}\,.
\ee
The different $N$ dependence of the two pieces comes from the fact that the first term in (\ref{G02222}) corresponds to the disconnected part of the correlator while the second term is the connected part.

\section{The operator product expansion}
\label{sec3}
We will consider the operator product expansion obtained in the limit $x_{12}^2 \rightarrow 0$, $x_{34}^2 \rightarrow 0$. This expansion of the four-point correlators has been extensively discussed in many papers \cite{Dolan:2000ut,hep-th/0112251,Dolan:2004iy,hep-th/0601148}. In cross-ratio variables it corresponds to the limit $u \rightarrow 0$ with $v$ fixed. The expansion of the correlator is then dictated by exchanged operators of a given twist (i.e. dimension minus spin), with the dominant terms given by the operators of lowest twist. The OPE is convergent and therefore if we keep all terms in the expansion (as we do in the following discussion) it is valid for all values of $u$ and $v$ inside the radius of convergence. In the following we often use the label $t$ to mean half the twist,
\be
t = \tfrac{1}{2}(\Delta - l)\,.
\ee

We will employ the superconformal blocks of \cite{1508.03611} which allow us to explicitly decompose the correlators into contributions from protected superconformal multiplets and unprotected ones. The correlation function $\langle{2222}\rangle$ then has the following OPE expansion
\begin{align}
\label{G2222OPE}
\langle{2222}\rangle = (g_{12}g_{34})^2 &+  (g_{12}g_{34}g_{13}g_{24}) A_{2,0} F^{\rm half}_{1,0} + (g_{13}g_{24})^2 A_{4,0} F^{\rm half}_{2,0} \notag \\
&+(g_{13}g_{24})^2 \biggl[ \sum_{l \geq 2} A_{4,l} F^{\rm sh.}_{2,l} + \sum_{l \geq 0} A'_{4,l} F^{\prime\, \rm sh.}_{2,l} \biggr] \notag \\
&+(g_{13}g_{24})^2 \sum_{t , l\geq0} C_{t,l} F^{\rm long}_{t,l} \,.
\end{align}
In the above decomposition the terms in the first line correspond respectively to the contributions of the identity operator, the half-BPS energy-momentum multiplet and a twist four half-BPS contribution. The terms in the second line with $l \geq 2$ comprise the contributions of the semi-short multiplets with primaries of twist four and spin $l$ in the $su(4)$ representation $[0,2,0]$ (for $F^{\rm sh.}_{2,l}$) or the $[1,0,1]$ representation (for $F^{\prime\, \rm sh.}_{2,l}$). The block $F^{\prime\, \rm sh.}_{2,0}$ gives the contribution of quarter-BPS multiplet whose primary has spin zero, twist four and $su(4)$ labels $[2,0,2]$. Finally the third line comprises the contributions of all long superconformal multiplets with twist $2t$ and spin $l$. In all terms the sum over $l$ is only over even spins, due to the symmetry of the correlator under the exchange of the first two operators.

The dependence on the Yang-Mills coupling enters only through the contributions of the long multiplets. The dimensions (and therefore the twists) of such multiplets are coupling dependent and hence generically not integer valued. Likewise the OPE coefficients $C_{t,l}$ are also  explicitly dependent on the coupling.

We give the explicit forms of the superconformal blocks and OPE coefficients for the protected multiplets in Appendix \ref{App-prot}. Our focus here is the contribution of the long multiplets. These are given by
\be
\label{longblocks2222}
F^{\rm long}_{t,l} = (x-y)(x-\bar{y})(\bar{x}-y)(\bar{x}-\bar{y}) G_{t,l}(x,\bar{x})\,,
\ee
where
\be
G_{t,l}(x,\bar{x}) = \frac{f_{t+l}(x) f_{t-1}(\bar{x}) - f_{t+l}(\bar{x}) f_{t-1}(x)}{{x-\bar{x}}}
\ee
and
\be
f_\rho(x) = x^{\rho-1} {}_2F_1(\rho+2,\rho+2,2\rho+4;x)\,.
\ee
The presence of the explicit factor of $(x-y)(x-\bar{y})(\bar{x}-y)(\bar{x}-\bar{y})$ in the blocks for long multiplets agrees with the expectation that all quantum corrections appear with such a prefactor in (\ref{2222int}) in accordance with partial non-renormalisation.

In free field theory (where $t$ is an integer) the coefficients $C_{t,l}$ take the following form for $t\geq2$,
\be
\label{freeCtl}
C_{t,l} = \frac{2(t+l+1)!^2 t!^2\bigl((l+1)(2t+l+2) + 4a(-1)^t\bigr)}{(2t)!(2t+2l+2)!}\,,
\ee
while for twist two ($t=1$) we have
\be
\label{twist2longcfs}
C_{1,l} = \frac{8a (l+2)!^2}{(2l+4)!}\,.
\ee
Recall that $l$ is to be taken even in these formulae.

\section{The supergravity limit and double-trace spectrum}
\label{sec4}

Here our primary focus is on the supergravity limit of $\mathcal{N}=4$ super Yang-Mills theory. This is a limit where we fix $g_{\rm YM}$ and take large $N$ and perform an expansion in $1/N^2$. In such a limit the 't Hooft coupling $\lambda = g^2_{\rm YM} N$ becomes large and operators dual to excited string states decouple as they become infinitely massive.

The protected single-trace half-BPS operators however are present in the spectrum. The energy-momentum multiplet corresponds to the graviton multiplet and the higher-charge half-BPS operators correspond to higher Kaluza-Klein modes from the reduction on $S^5$.

Similarly operators built from products of single-trace half-BPS operators are also predicted to remain in the spectrum. Such operators can themselves be protected or they can be unprotected. The unprotected operators of this type are still present in the spectrum because in the strictly infinite $N$ limit they keep their classical scaling dimensions due to operator factorisation, and hence the corresponding states do not acquire infinite mass. In the supergravity spectrum such operators are `nearly' protected and receive anomalous dimensions at order $1/N^2$ and higher.

However, all other operators, not built from products of singe-trace half-BPS operators, correspond to the afore-mentioned string states. Such operators are therefore absent from the spectrum in the supergravity limit. All of the twist-two long operators in the expansion (\ref{G2222OPE}) are of this type.

The simplest long operators which remain in the supergravity spectrum are double-trace operators and we will examine their spectrum by analysing the four-point functions of the single-trace half-BPS operators. The double-trace operators $K_{p,q}^{t,l} = \mathcal{O}_{p}\Box^n \partial^l \mathcal{O}_{q}$ are special for two reasons. Firstly we expect the three-point functions $\langle \mathcal{O}_{p'} \mathcal{O}_{q'} K_{p,q}^{t,l}\rangle$ to be non-zero already at the leading order in the $1/N^2$ expansion, whereas we expect the three-point functions involving triple-trace operators and higher to be suppressed. Secondly, there is a unique operator of the form $K_{p,q}^{t,l}$ of spin $l$ for fixed $p,q,t$ and fixed $su(4)$ labels. The triple-trace and higher multi-trace operators do not obey this property; their number grows with the spin.

For the correlation function of particular interest here, $\langle{2222}\rangle $ it is convenient to change the expansion parameter from $1/N^2$ to 
\be
a= \frac{1}{N^2-1}\,
\ee
so that we have
\be
\langle{2222}\rangle =  \sum_{n=0}^{\infty} a^n \langle 2222 \rangle^{(n)}\, .
\ee
This has the benefit that the free theory correlator then contributes to just the first two terms. The  interacting part of the correlator and hence the function $F(u,v)$ appearing there have an expansions of the form
\be
\langle 2222 \rangle_{\rm int} = \sum_{n=1}^\infty a^n \langle 2222 \rangle_{\rm int}^{(n)}\,, \qquad F(u,v) = \sum_{n=1}^{\infty} a^n F^{(n)}(u,v)\,,
\ee
so that they contribute to all terms except the leading one. In other words we have
\begin{align}
\langle 2222 \rangle^{(0)}&= \langle 2222 \rangle^{(0)}_{\rm free} \,, \notag \\
\langle 2222 \rangle^{(1)} &= \langle 2222 \rangle^{(1)}_{\rm free} + \langle 2222 \rangle^{(1)}_{\rm int} \,,
\end{align}
while for $n\geq 2$ we have
\be
\langle 2222 \rangle^{(n)}= \langle 2222 \rangle^{(n)}_{\rm int} = g_{13}^2 g_{24}^2 s(x,\bar x;y,\bar y)  F^{(n)}(u,v)\,.
\ee
Here $\langle 2222 \rangle^{(0)}$ corresponds to the contribution of disconnected supergravity diagrams. The connected tree-level Witten diagrams contribute to $\langle 2222 \rangle^{(1)}$, while $\langle 2222 \rangle^{(2)}$ corresponds to one-loop supergravity corrections. From tree-level supergravity we have \cite{hep-th/9911222,hep-th/0002170}
\be
F^{(1)}(u,v) = - 4 \partial_u \partial_v(1+u \partial_u + v \partial_v) \Phi^{(1)}(u,v) = -4 \bar{D}_{2422}(u,v)\,,
\ee
where $\Phi^{(1)}(u,v)$ is the one-loop scalar box integral and we also give the expression in terms of the $\bar{D}$-functions introduced in \cite{hep-th/0112251}.

Our task now is to compare the known result with the general form of the operator product expansion given in eq. (\ref{G2222OPE}). We are interested in the contribution of the long multiplets. At the leading order in the $a$ expansion the only contributions are long operators of twist four and higher, corresponding to the the term proportional to $A$ in the free-field expression (\ref{freeCtl}). This encompasses the contributions of all the long double-trace operators to the disconnected part of the correlator. For the $\langle 2222 \rangle$ correlation function, there is only one $su(4)$ channel; all the superconformal primaries are in the singlet. We must remember that at leading order all these operators will have their classical scaling dimensions (and hence twists). This means that many operators can be degenerate and the twist is not a good label for the spectrum. Thus we introduce a new label $i$ to run over different operators which share the same quantum numbers at leading order in $a$. In fact we may count the number of such operators simply: at twist $2t$ there are $(t-1)$ such degenerate operators for each spin $l$,
\be
\label{doubletraceops}
\{(\mathcal{O}_{2} \Box^{t-2} \partial^l \mathcal{O}_{2}), (\mathcal{O}_{3} \Box^{t-3} \partial^l \mathcal{O}_{3}), \ldots, (\mathcal{O}_{t} \Box^0 \partial^l \mathcal{O}_{t})\}\,.
\ee
We denote such double-trace operators by $K_{t,l,i}$ for $i=1,\ldots,t-1$. Thus we have a relation for the three-point functions at leading order
\be
\label{leading3ptfns}
\sum_{i=1}^{t-1} \langle \mathcal{O}_{2} \mathcal{O}_{2} K_{t,l,i} \rangle^2 = C^{(0)}_{t,l}\,.
\ee
where the $C^{(0)}_{t,l}$ are given by the coefficients in (\ref{freeCtl}) with $a$ set to zero.
\be
C^{(0)}_{t,l} =  \frac{2(t+l+1)!^2 t!^2 (l+1)(2t+l+2)}{(2t)!(2t+2l+2)!}
\ee
The relation (\ref{leading3ptfns}) holds under the assumptions outlined earlier that only the operators listed in (\ref{doubletraceops}) contribute at leading order in the $ a$ expansion.

When we proceed to the next order in $a$ there are many issues to take into account. Firstly, all the twist-two long multiplets which are present in free field theory must be absent in the supergravity spectrum. It is therefore necessary that the contribution of such multiplets to $\langle 2222 \rangle^{(1)}_{\rm int}$ cancels the contribution from the connected part of the free theory correlation function $\langle 2222 \rangle_{\rm free}^{(1)}$ which corresponds to the coefficients given in eq. (\ref{twist2longcfs}). This is indeed the case \cite{hep-th/0112251}. The only twist-two contribution in supergravity is therefore the protected energy-momentum multiplet, which is the dual of the graviton multiplet.

Next, at the first subleading order we must take into account the fact that the long double-trace operators develop anomalous dimensions. The true twist of the operator $K_{t,l,i}$ is therefore no longer $2t$ (which we still take to be integer) but rather it is
\be
2(t +  a \eta_{t,l,i}^{(1)} + a^2 \eta_{t.l,i}^{(2)} + \ldots)\,.
\ee
Here we use the notation that $\eta_{t,l,i} = \sum_{n=1}^{\infty} a^n \eta^{(n)}_{t,l,i}$ is half the anomalous dimension of the operator $K_{t,l,i}$. Examining the expression for the blocks given in (\ref{longblocks2222}) and performing the perturbative expansion in $a$ we find that the functions $F^{(n)}(u,v)$ will be expressible as a series of logarithms with coefficients which are analytic functions of $u$,
\be
\label{Fnlogdecomp}
F^{(n)}(u,v) = \sum_{r=0}^{n} (\log u)^r F^{(n)}_{r}(u,v)\,.
\ee
At order $a$ we find a contribution to the OPE proportional to $\log u$,
\be
\label{logterm}
F^{(1)}_1 (u,v) =\sum_{t=2}^{\infty} \sum_{l=0}^{\infty} \sum_{i=1}^{t-1} \langle \mathcal{O}_{2} \mathcal{O}_{2} K_{t,l,i} \rangle^2 \eta_{l,i}^{(1)} G_{t,l}(x,\bar{x})\,.
\ee
The contribution of such terms in the OPE must match the part of the explicit result for $F^{(1)}$ with a logarithmic branch cut around $u=0$ (i.e. the discontinuity around $u=0$). Decomposing $F^{(1)}(u,v)$ into a series of logarithms with analytic coefficients as in (\ref{Fnlogdecomp}) we find
\be
\label{F1logs}
F_1^{(1)}(u,v) = \frac{8}{v(x-\bar{x})^6}\biggl(f(u,v) + v g(u,v) \frac{{\rm Li}_1(x) - {\rm Li}_1(\bar{x})}{x-\bar{x}}\biggr)\,,
\ee
where
\begin{align}
f(u,v) = \,& -1 + 3 u - 3 u^2 + u^3 - 8 v - 9 u v + 16 u^2 v + u^3 v  \notag \\
&+ 18 v^2 - 9 u v^2 - 3 u^2 v^2 - 8 v^3 + 3 u v^3 - v^4 \notag \\
g(u,v) =\, & 6(1 - u - u^2 + u^3 - v + 4 u v - u^2 v - v^2 - u v^2 + v^3)\,.
\end{align}
Identifying the expressions (\ref{logterm}) and (\ref{F1logs}) one finds \cite{hep-th/0112251} 
\be
\label{etax3ptfns}
\sum_{i=1}^{t-1}  \langle \mathcal{O}_{2} \mathcal{O}_{2} K_{t,l,i} \rangle^2 \eta_{t,l,i}^{(1)}= -C^{(0)}_{t,l}\frac{(t-1)t(t+1)(t+2)}{2(l+1)(2t+l+2)} \,.
\ee

Due to the sum over $i$ in (\ref{leading3ptfns}) and (\ref{etax3ptfns}) one cannot generically deduce the anomalous dimensions $\eta_{l,i}^{(1)}$ and leading order three-point functions $\langle \mathcal{O}_{2} \mathcal{O}_{2} K_{t,l,i} \rangle$. This phenomenon is known as operator mixing. The exception is the case of twist four $(t=2)$ where there is only a single double-trace operator $K_{2,l}$ for each spin $l$ and hence no mixing. In this case one finds simply that \cite{hep-th/0112251} 
\be
\eta^{(1)}_{2,l} = - \frac{48}{(l+1)(l+6)}\,, \qquad \langle \mathcal{O}_{2} \mathcal{O}_{2} K_{2,l} \rangle^2 =  \frac{(l+3)!^2 (l+1)(l+6)}{3(2l+6)!}\,.
\ee

To proceed further we make use of the fact that all correlators of the form $\langle p_1 p_2 p_3 p_4 \rangle$ are known due to the work of Rastelli and Zhou \cite{1608.06624}. The formula found in \cite{1608.06624} is consistent with many previously known cases found by other methods \cite{hep-th/0002170,hep-th/0601148,0811.2320,1106.0630}. We develop a very similar OPE analysis for all correlators, both for the large $N$ free field expressions and for the $\log u$ terms found from the results of \cite{1608.06624}. In particular analysing the singlet channel of the correlators of the form $\langle ppqq \rangle$ provides us with enough information to deduce all the three-point functions $\langle \mathcal{O}_{p} \mathcal{O}_{p} K_{t,l,i}\rangle$. Indeed one may go further and deduce similar information for the non-singlet channels as well. We will provide much more information on this analysis in a forthcoming paper \cite{Us-unmixing}. Here we simply quote the results of relevance to the study of the $\langle 2222 \rangle$ correlation function. We find for the anomalous dimensions,
\be
\eta_{t,l,i} = -\frac{2(t-1)_4 (l+t)_{4}}{(l+2i-1)_{6}}\,,
\ee
where $(x)_n = (x+n-1)!/(x-1)!$ denotes the Pochhammer symbol.
For the three-point functions we find 
\be
\langle \mathcal{O}_{2} \mathcal{O}_{2} K_{t,l,i} \rangle^2 = C^{(0)}_{t,l} R_{t,l,i} a_{t,i}
\ee
with
\begin{align}
R_{t,l,i} &= \frac{2^{1-t}(2l+3+4i) (l+i+1)_{t-i-1} (t+l+4)_{i-1}}{(\tfrac{5}{2}+l+i)_{t-1}}\,,\notag \\
a_{t,i} &= \frac{2^{(1 - t)} (2 + 2 i)! (t-2)! (2t - 2 i + 2)!}{3 (i-1)! (i+1)! (t+2)! (t-i-1)! (t-i+1)!}\,.\label{3pnt}
\end{align}
The fact that the results for the both the anomalous dimensions and the three-point functions are so simple and in a closed form is already something of a miracle. Note, for example, that the spin dependence always factorises completely into linear factors in $l$ in both quantities. As far as we are aware these are the first results on double-trace anomalous dimensions and three-point functions for arbitrary twist and spin. We will now use these results to make predictions for the one-loop supergravity corrections to the correlation function, i.e. the function $F^{(2)}(u,v)$.

\section{Resummation of the one-loop double discontinuity}
\label{sec5}
With the results for the three-point functions and anomalous dimensions to hand we can now make a prediction for the leading $\log^2 u$ term in $F^{(2)}(u,v)$, in other words, the coefficient $F_2^{(2)}(u,v)$ in the expansion (\ref{Fnlogdecomp}). From expanding the blocks to order $a^2$ we find a contribution to the OPE with a double logarithm of $u$ and the square of the anomalous dimensions. Thus at order $a^2$ we predict a double discontinuity contribution to $F^{(2)}(u,v)$ of the form
\be
F_2^{(2)}(u,v) = \frac{1}{2}   \sum_{t=2}^{\infty} \sum_{l=0}^{\infty} \sum_{i=1}^{t-1}  \langle \mathcal{O}^{(2)} \mathcal{O}^{(2)} K_{t,l,i} \rangle^2  (\eta_{t,l,i}^{(1)})^2 G_{t,l}(x,\bar{x})\,.
\ee
We perform the sum by obtaining many orders in $x$ and $\bar{x}$ variables. Comparing the series to a plausible ansatz in terms of polylogarithmic functions we find the following form for the double discontinuity,
\begin{align}
\label{dbldisc}
F_2^{(2)}(u,v) = \frac1{uv}\Big[&p(u,v) \frac{{\rm Li}_1(x)^2 - {\rm Li}_1(\bar{x})^2}{x-\bar{x}} + 2\biggl[p(u,v)+p\biggl(\frac{1}{v},\frac{u}{v}\biggr)\biggr]\frac{{\rm Li}_2(x) - {\rm Li}_2(\bar{x})}{x-\bar{x}} \notag \\
& + q(u,v) ({\rm Li}_1(x) + {\rm Li}_1(\bar{x})) + r(u,v)\frac{{\rm Li}_1(x) - {\rm Li}_1(\bar{x})}{x-\bar{x}} + s(u,v)\Big]\,.
\end{align}
where $p,q,r,s$ are rational functions of $u$ and $v$. We may then check the obtained result to very high orders in both variables, finding perfect agreement.

The coefficient function $p$ is symmetric $p(u,v) = p(v,u)$ as required by crossing since the double discontinuity in both $u$ and $v$ comes only from the first term in (\ref{dbldisc}) which contributes $p(u,v) \log^2 u \log^2 v$ and hence must be symmetric in $u$ and $v$. As we will see, the fact that the coefficient of the ${\rm Li}_2$ term is related simply to the same function $p$ is a hint at an additional simplicity in the final amplitude.

It is possible to write the coefficient $p(u,v)$ in quite a simple form, 
\be
p(u,v) = 24 u v \partial_x^2 \partial_{\bar{x}}^2 \biggl[\frac{u^2 v^2 (1-u-v)[(1-u-v)^4+20 u v (1-u-v)^2 + 30 u^2 v^2]}{(x-\bar{x})^{10}}\biggr]\,.
\ee
The other coefficients are more complicated and we will not give their explicit expressions. Instead we will proceed to construct a fully crossing symmetric function $F^{(2)}(u,v)$ with the correct double discontinuity. The remaining coefficients in (\ref{dbldisc}) can then be obtained from $F^{(2)}(u,v)$ by taking the double discontinuity.

\section{Completion to a crossing symmetric amplitude}
\label{sec6}
Having obtained the double discontinuity from resumming the OPE, we make an ansatz for the form of the full crossing invariant contribution to supergravity at one loop. In order to construct a suitable ansatz we note that the tree-level supergravity function $F^{(1)}(u,v)$ is expressible in terms of a $\bar{D}$-function which is a particular combination of derivatives acting on the one-loop box function $\Phi^{(1)}(u,v)$. This means that it is expressible as a combination of single-valued polylogarithms of weights 2,1 and 0 with rational functions of $x$ and $\bar{x}$ as coefficients. The particular class of single-valued polylogarithms of interest here are linear combinations of polylogarithms constructed on the singularities (or `letters') $\{x,1-x, \bar{x}, 1- \bar{x}\}$ such that they are single-valued when $\bar{x}$ is taken to be the complex conjugate of $x$. They are constructed in general in \cite{fbsvpl} and appear in many contexts to discuss the perturbative contributions to the correlation functions $\langle p_1 p_2 p_3 p_4 \rangle$ \cite{1303.6909,1512.02926} as well as in multi-Regge kinematics of scattering amplitudes \cite{Dixon:2012yy,DelDuca:2016lad} and Feynman integral calculations \cite{Drummond:2012bg,Schnetz:2013hqa}.

Since our result for the double discontinuity $F_2^{(2)}(u,v)$ given in eq. (\ref{dbldisc}) is expressible in terms of logarithms and dilogarithms it seems a natural choice is to construct an ansatz for the full function $F^{(2)}(u,v)$ from the same class of single-valued polylogarithms, but this time of weights 4,3,2,1, and 0 with rational functions for coefficients. We then impose crossing symmetry and the fact that the double discontinuity must match our result for $F_2^{(2)}(u,v)$. 

The constraints described in the previous paragraph fix completely the weight 4 and weight 3 parts of the result with rational coefficients which are determined by the coefficients appearing in $F_2^{(2)}(u,v)$. The weight 2, 1 and 0 parts are not fixed completely by matching to the double discontinuity.
Since the double discontinuity $F_2^{(2)}(u,v)$ has 15 powers of $(x-\bar{x})$ in the denominator so do the rational coefficients in the weight 4 and weight 3 parts. This leaves the possibility that the resulting function has unphysical poles at $x=\bar{x}$. 

In order to make sure that poles at $x=\bar{x}$ are in fact absent, we have to arrange the weight 2,1 and 0 parts so that they cancel those of the weight 4 and weight 3 pieces. We then allow a maximum of 15 powers of $(x-\bar{x})$ in the denominators of the coefficients of the weight 2,1 and 0 parts of the ansatz to match the denominators in the weight 4 and weight 3 parts and demand that all poles at $x=\bar{x}$ cancel. We also demand that the twist-two sector is completely absent\footnote{Recall the twist-two long operators are absent from the supergravity spectrum and the cancellation of such contributions between $\langle 2222 \rangle_{\rm free}^{(1)}$ and $\langle 2222 \rangle^{(1)}_{\rm int}$ is complete. Therefore there should be no twist-two contributions in $F^{(2)}$.} from $F^{(2)}(u,v)$. These constraints completely fix the answer within our ansatz up to a single free coefficient. 

We find we can express the final crossing symmetric result in terms of ladder integrals \cite{Usyukina:1992jd,Isaev:2003tk}. These are a particular subset of the single-valued polylogarithms under considerations here. They are given by
\be
\Phi^{(l)}(u,v) = - \frac{1}{x-\bar{x}} \phi^{(l)}\biggl(\frac{x}{x-1},\frac{\bar{x}}{\bar{x}-1}\biggr)\,,
\ee
where
\be
\phi^{(l)}(x,\bar{x}) = \sum_{r=0}^l (-1)^r \frac{(2l-r)!}{r!(l-r)! l!} \log^r (x \bar{x}) ({\rm Li}_{2l-r}(x) - {\rm Li}_{2l-r}(\bar{x}))\,.
\ee
The functions $\Phi^{(l)}$ obey
\be
\Phi^{(l)}(u,v) = \Phi^{(l)}(v,u)
\ee
while $\Phi^{(1)}$ also obeys
\be
\frac{1}{u}\Phi^{(1)}\biggl(\frac{1}{u},\frac{v}{u}\biggr) = \Phi^{(1)}(u,v)\,.
\ee

We recall that the correlation function in the supergravity limit then takes the form~\eqref{G2222genform}
\be
\langle{2222}\rangle = \langle{2222}\rangle_{\rm free} + g_{13}^2 g_{24}^2 s(x,\bar x;y,\bar y) F(u,v)\,.
\ee
with
\be
F(u,v)  = a F^{(1)}(u,v) + a^2 F^{(2)}(u,v) +O(a^3)\,.
\ee
The tree-level supergravity contribution is given by
\be
F^{(1)}(u,v) = -4 \partial_u  \partial_v(1 + u \partial_u + v \partial_v)  \Phi^{(1)}(u,v)\,.
\ee

Our final result for the one-loop correction contains a single unfixed parameter within the ansatz outlined above. We first quote a particular solution where we set the free parameter $\alpha$ to zero. Then we will give the ambiguity. In the next section we will argue that $\alpha=0$ is in fact needed to maintain analyticity in the spin for the twist-four anomalous dimensions at order $a^2$.

Our particular solution is given by the crossing symmetric combination
\be
F^{(2)}(u,v) = \frac{1}{uv}\Big[f(u,v) + \frac{1}{u}f\biggl(\frac{1}{u},\frac{v}{u}\biggr) + \frac{1}{v} f\biggl(\frac{1}{v},\frac{u}{v}\biggr)\Big]\,.
\label{forientations}
\ee
To simplify the presentation of the function $f(u,v)$ we write
\be
f(u,v) = \Delta^{(4)} g(u,v)\,, \qquad \Delta^{(4)} = (x-\bar{x})^{-1} u v \partial_x^2 \partial_{\bar{x}}^2 (x-\bar{x})\,.
\ee
Furthermore we can decompose the function $g$ into pieces according to the transcendental weight of the polylogarithmic contributions
\be
g = (x-\bar{x})^{-10}[g^{(4)}+g^{(3)}+g^{(2)}+g^{(1)}+g^{(0)}]\,.
\ee
The pieces of given weight are then as follows,
\begin{align}
\label{loopamplitude}
g^{(4)}(u,v) =& P_-^{(4)}(u,v) \Phi^{(2)}(u,v) \notag \\
g^{(3)}(u,v) =& P^{(3)}_+(u,v) \Psi(u,v) + P^{(3)}_-(u,v) \log (u v)\Phi^{(1)}(u,v) \notag \\
g^{(2)}(u,v) =& P^{(2)}_+(u,v) \log u \log v + P^{(2)}_-(u,v) \Phi^{(1)}(u,v) \notag \\
g^{(1)}(u,v) =& P_+^{(1)}(u,v) \log (u v)  \notag \\
g^{(0)}(u,v) =& P_+^{(0)}(u,v)\,.
\end{align}
The function $\Psi(u,v)$ is a particular derivative of the two-loop ladder integral,
\begin{align}
\Psi(u,v) &= (x-\bar{x})(u \partial_u + v \partial_v)[(x-\bar{x})\Phi^{(2)}(u,v)] \notag \\
&= [x(1-x)\partial_x - \bar{x}(1-\bar{x})\partial_{\bar{x}}]\phi^{(2)}\biggl(\frac{x}{x-1},\frac{\bar{x}}{\bar{x}-1}\biggr)\,.
\end{align}
The coefficients $P_{\pm}^{(r)}(u,v)$ in (\ref{loopamplitude}) are symmetric polynomials in $u$ and $v$. The subscripts $\pm$ correspond to the symmetry properties under $x \leftrightarrow \bar{x}$ of the pure transcendental factor that each coefficient $P^{(r)}$ multiplies (antisymmetric for the ladder functions and symmetric for constants, for logarithms of $u$ and $v$ and for $\Psi(u,v)$) . Note that the weight four piece is entirely expressible in terms of $\Phi^{(2)}(u,v)$, whose transcendental part is antisymmetric in $x$ and $\bar{x}$. In principle there could have been a symmetric part, e.g. $\Phi^{(1)}(u,v)^2$, but in fact our function does not have such a contribution. The fact that the weight four piece is given by $\Phi^{(2)}(u,v)$ only implies the relationship between the coefficients of the ${\rm Li}_2$ terms and the ${\rm Li}_1^2$ terms in the double discontinuity (\ref{dbldisc}).

To express the coefficient polynomials it is helpful to introduce symmetric variables
\be
\bar{s} = 1- u -v\,, \qquad p = u v\,.
\ee
The coefficient polynomials are then given by
\begin{align}
P^{(4)}_-(u,v) &= 96 p^2 \bar{s}[\bar{s}^4+20 p \bar{s}^2 + 30 p^2]\,, \\
P^{(3)}_+(u,v) &= \frac{8}{5} p^2 [137 \bar{s}^4 + 1214 p \bar{s}^2 + 512 p^2]\,, \\
P^{(3)}_-(u,v) &= 336 p^2 [\bar{s} (1-\bar{s})  (6 - 6\bar{s} + \bar{s}^2)
 +2 p(3 - 14 \bar{s} + 4 \bar{s}^2)  - 16 p^2]\,,\\
P^{(2)}_+(u,v) &= 2[(1 - \bar{s})^2 \bar{s}^6 -2 p \bar{s}^4 (20 - 33 \bar{s} + 14 \bar{s}^2) \notag \\& \qquad + 8 p^2 (756 - 
   1323 \bar{s} + 601 \bar{s}^2 - 54 \bar{s}^3 + 30 \bar{s}^4) \notag \\
   & \qquad -32 p^3 (583 - 25 \bar{s} + 
   26 \bar{s}^2)+ 1024 p^4]\,, \\
P^{(2)}_-(u,v) &=  56 p^2[-\bar{s}^2(2 - \bar{s})  (18 - 18 \bar{s} + 5 \bar{s}^2) \notag \\
&\qquad \qquad +2 p(108 - 144 \bar{s} + 128 \bar{s}^2 - 
   11 \bar{s}^3)- 8 p^2 (63 - \bar{s})]\,,\\
P^{(1)}_+(u,v) &=  \frac{1}{3}[5 \bar{s}^7 (2 - 3 \bar{s})-
 2 p \bar{s}^5 (158 - 193 \bar{s}) \notag \\
 &\qquad +16 p^2 \bar{s} (378 - 567 \bar{s} + 233 \bar{s}^2 - 
    147 \bar{s}^3) \notag \\
    & \qquad + 32 p^3 (378 - 139 \bar{s} + 129 \bar{s}^2) + 256 p^4] \, ,\\
 P^{(0)}_+(u,v) & = \frac{2}{15}(x-\bar{x})^2[20 (1 - \bar{s}) \bar{s}^6 - 
 5 p \bar{s}^4 (102 - 75 \bar{s} - 4 \bar{s}^2) \notag \\
 &\qquad \qquad \qquad +8 p^2 (630 - 630 \bar{s} + 481 \bar{s}^2 - 
    255 \bar{s}^3 - 30 \bar{s}^4) \notag \\
    &\qquad \qquad \qquad -16 p^3 (217 - 215 \bar{s} - 60 \bar{s}^2) -1280 p^4]\,.
    \label{P0}
\end{align}

The terms involving $P^{(4)}_-,P^{(3)}_\pm,P^{(2)}_+$ contribute to the double discontinuity and therefore the coefficients are related to those appearing in (\ref{dbldisc}). In particular we have
\be
p(u,v) = \frac{1}{4} (x-\bar{x}) \Delta^{(4)} \biggl[\frac{P^{(4)}_-(u,v)}{(x-\bar{x})^{11}}\biggr]\,.
\ee

The ambiguity in the result is much simpler. In fact all terms proportional to the single free parameter $\alpha$ can be expressed in a similar way to the tree-level amplitude,
\be
\alpha \frac{1}{uv}[(1 + u \partial_u + v \partial_v) u \partial_u v \partial_v]^2  \Phi^{(1)}(u,v)\,.
\label{ambig}
\ee
At this stage our solution is given by the particular solution $F^{(2)}(u,v)$, as described in equations (\ref{forientations})-(\ref{P0}), plus the amibiguity in eq. (\ref{ambig}) above. Note that the ambiguity in (\ref{ambig}) has no double discontinuity, has no unphysical poles, is fully crossing symmetric and has no twist-two contribution. When written out in terms of single-valued polylogarithms with rational coefficients, the ambiguity in (\ref{ambig}) has 13 powers of $(x-\bar{x})$ in the denominator. In terms of $\bar{D}$-functions it can be expressed as $\bar{D}_{4444}$. In the next section we argue that $\alpha=0$.

We should sound a note of caution that what we have presented is not strictly a derivation of the one-loop correction. It is possible that the true answer differs from the expression we have constructed above by a function that itself has no double discontinuity, no unphysical poles, no twist-two sector and is fully crossing symmetric on its own.

In principle there are further ambiguities we could add within the class of single-valued polylogarithms multiplied by rational functions. These all have higher powers of $(x-\bar{x})$ in the denominator than the 15 we allowed above. They correspond to crossing symmetric $\bar{D}$-functions with higher weights. Indeed such functions have arisen in the context of possible stringy corrections \cite{1410.4717}. 

Finally, it is also possible that there are functions which do not sit in the class of single-valued polylogarithms that we have allowed. However it is highly non-trivial that we are able to find a solution, unique up a single free parameter within the simplest class of functions we are led to consider and we take this as very strong encouragement that our amplitude is in fact correct.

While the result presented above in equations (\ref{forientations})-(\ref{P0}) and (\ref{ambig}) is certainly one way to represent the result of our crossing symmetric one-loop amplitude, we do not claim that it is necessarily the most natural. It seems highly likely that it will be simpler in its Mellin space representation, as the tree-level result is for general charges \cite{1608.06624}. 

\section{Twist 4 anomalous dimensions at order $a^2$}
\label{sec7}
Having obtained the correlation function at NNLO we can try to extract anomalous dimensions of the double trace operators from it. These should correspond to loop corrections to the masses of the corresponding multi-particle supergravity states via AdS/CFT. The order $a^2$ anomalous dimensions appear within the partial wave decomposition of the single discontinuity of the correlation function we have constructed in the previous section. For general external half-BPS operators and exchanged operators of general twist, we expect triple-trace operators to also contribute to the single discontinuity at order $a^2$, although they are absent from the double discontinuity used to construct the correlator. However at twist four there are no such triple-trace operators and furthermore there is a single double-trace operator for each spin so we can extract the anomalous dimensions of such double-trace operators as we will now show.

Recall that the correlator takes the form
\begin{align}
	\langle 2222  \rangle =&~ \langle 2222  \rangle_{\text{free}}^{(0)} + a \langle 2222  \rangle_{\text{free}}^{(1)} \notag \\ &+ a \langle 2222  \rangle^{(1)}_{\text{int}} + a^2 \langle 2222  \rangle^{(2)}+ \mathcal{O}(a^3)\label{cor2222}
\end{align}
where 
\begin{align}
	\langle 2222  \rangle_{\text{free}}^{(0)} &= \Big(g_{12}^2g_{34}^2+g_{13}^2g_{24}^2+g_{14}^2g_{23}^2\Big)\\
		\langle 2222  \rangle_{\text{free}}^{(1)}&= 4 \Big(g_{12}g_{34}g_{13}g_{24}+g_{12}g_{34}g_{14}g_{23}+g_{13}g_{24}g_{14}g_{23}\Big)\\
		\langle 2222  \rangle^{(1)}_{\text{int}} &= -4\,g_{13}^2 g_{24}^2 s(x,\bar{x};y,\bar{y}) \bar D_{2422}(x,\bar x)\\
		\langle 2222  \rangle^{(2)} &= 4\,g_{13}^2 g_{24}^2 s(x,\bar{x};y,\bar{y}) F^{(2)}(u,v)
		\end{align}
		and $F^{(2)}(u,v)$ is our new result given in~\eqref{forientations}.
We wish to equate this to a superconformal partial wave expansion.

Focussing on the twist-four operators, the SCPW expansion reads
\begin{align}
\langle 2222  \rangle|_{\text{twist 4 sector}} = g_{13}^2g_{24}^2 \sum_{l\in2\mathbb{N}} C_{2,l}(N) F^{\text{long}}_{2+\gamma,l} \label{t4}
\end{align}
where 
\begin{align}
	F^{\text{long}}_{2+\gamma,l}= s(x,\bar{x};y,\bar{y})(x\bar{x})^{\frac{\gamma}{2}} G_{2+\gamma,l}(x,\bar{x})
\end{align}
and where 
\begin{align}
  G_{2+\gamma,l}(x,\bar{x}) = \frac{x^{l+1} {}_2F_1(4{+}l{+}\frac \gamma 2,4{+}l{+}\frac \gamma 2,8{+}2l{+}\gamma,x) {}_2F_1(3{+}\frac \gamma 2,3{+}\frac \gamma 2,6{+}\gamma,\bar x) - x\leftrightarrow \bar x}{x-\bar x}\ .
\end{align}
Since this discussion is focussed only on twist four, we define for convenience
\begin{align}
A_l:=C_{t=2,l}.
\end{align}
Expanding the normalisation and anomalous dimension in $N$ as:
\begin{align}
	A_l(N)&= A^{(0)}_l +a A^{(1)}_l+a^2 A^{(2)}_l + \mathcal{O}(a^3) \notag \\
		\eta_l(N)&=a \eta^{(1)}_l+ a^2 \eta^{(2)}_l+  \mathcal{O}(a^2)
\end{align}
we get the following expansion of the SCPWs~\eqref{t4}
 \begin{align}
\langle2222\rangle|_{\text{twist 4 sector}}=& s(x,\bar{x};y,\bar{y})\sum_{l\in2\mathbb{N}} \Bigg[	A_l^{(0)}G_{2,l} \notag \\
&+   a \left( \log(x \bar x)  A_l^{(0)}\eta^{(1)}_l  G_{2,l}+ A_l^{(1)}G_{2,l} + A_l^{(0)} \eta^{(1)}_l\frac \partial {\partial \gamma} G_{2+\gamma,l} \right)\notag \\
&+a^2\Bigg(
\log^2(x \bar x)  \frac {A_l^{(0)}}2 \big(\eta^{(1)}_l\big)^2  G_{2,l} \notag\\& \qquad\quad + \log(x \bar x)  \Big( {A_l^{(1)}} \eta^{(1)}_l G_{2,l} +  A_l^{(0)} {\eta^{(2)}_l} G_{2,l} +A_l^{(0)} \big(\eta^{(1)}_l\big)^2 \frac \partial {\partial \gamma} G_{2+\gamma,l}\Big) \notag \\
&\qquad\quad+ \log^0(x \bar x)\Big( ...\Big)
\Bigg) ~ + ~\mathcal{O}(a^3)
       \Bigg],
 \end{align}
 
 We now equate this expansion to the correlator~\eqref{cor2222}. 
The problem can be split up into a number of separate pieces, separating out the free theory from the tree-level and one-loop interacting pieces. So we have
 \begin{align}
 	\langle 2222 \rangle_{\text{free}}^{(0)}|_\text{twist 4 sector} &= g_{13}^2 g_{24}^2 \sum_{l}A_l^{(0)} F^{\text{long}}_{2,l}\\
 	\langle 2222 \rangle_{\text{free}}^{(1)}|_\text{twist 4 sector} &= g_{13}^2 g_{24}^2 \sum_{l}A^{(1)}_{\text{free},l} F^{\text{long}}_{2,l}\ ,
 \end{align}
  where we split the order $a$ contribution to the normalisation into a piece arising from connected free theory and a piece from supergravity 
 \begin{align}
 A^{(1)}_l=A^{(1)}_{\text{free},l}+A^{(1)}_{\text{int},l}\ .
 \end{align}
  The conformal partial wave analysis of the free theory is well known and was first computed in~\cite{hep-th/0112251} and reproduced more recently in these conventions in~\cite{1508.03611}. It yields
\begin{align}
	A^{(0)}_l=\frac{ (l+1) (l+6) ((l+3)!)^2}{3 (2 l+6)!}, \qquad A^{(1)}_{\text{free},l}= \frac{4 ((l+3)!)^2}{3 (2 l+6)!}\ .
\end{align} 
Next we consider the supergravity contribution to the correlator, which one splits into the $\log x\bar x$ contribution and a $\log^0 x\bar x$ contribution. We equate
\begin{align}
-4\bar D_{2422}|_{\log(x \bar x)} &= A_l^{(0)}\eta^{(1)}_l  G_{2,l} +\mathcal{O}(x \bar x)\\
-4\bar D_{2422}|_{\log^0(x \bar x)} &= \text{twist 2 contribution} + A^{(1)}_{\text{int},l}G_{2,l} + A_l^{(0)} \eta^{(1)}_l  \frac \partial {\partial \gamma}  G_{2+\gamma,l} +\mathcal{O}(x \bar x)\ .\label{2nd}
\end{align}
The first equation yields~\cite{hep-th/0112251}
\begin{align}
A_l^{(0)}\eta^{(1)}_l =	-\frac{16 ((l+3)!)^2}{(2 l+6)!} \qquad \Rightarrow \qquad \eta^{(1)}_l =-\frac{48}{(l+1) (l+6)}\ .
\end{align}
Plugging these values into~\eqref{2nd}, and including the SCPW for the twist operators one first obtains a twist two sector precisely cancelling that from the free theory, thus reproducing the well-known result that the twist 2 sector drops out. Secondly we obtain values for the correction to the normalisation due to supergravity which we haven't found a closed formula for. The first 20 even spin terms are: 
{\scriptsize \begin{align*}
	\Big\{&A^{(1)}_{\text{int},0},A^{(1)}_{\text{int},2},A^{(1)}_{\text{int},4},\dots\Big\}=\\
	\Big\{
	&\frac{14}{75},\frac{367}{19845},\frac{57191}{38648610},\frac{407117}{3723149430},\frac{39792607}{5131853787060},\frac{4665834631}{8701123888957500},\frac{58812219091}{1612868735890269000},\notag\\
	&\frac{3685539014567}{1504379594743767670500},\frac{962595120061373}{5901478806725536789755000},\frac{39734035774806913}{3684714753239233013329029000},\notag\\
	&\frac{1718092704673177939}{2423610205875655200585311130000},\frac{567891105901482934553}{12242252973728717079772550894375400},\notag\\
	&\frac{30217552473152404437509}{9993236004349870542057754236907837200},\frac{30320562388695385278449}{154329038786388394824389455826138274000},\notag\\
	&\frac{109446992694061123595058871}{8597820765253205965031168033609718974647200},\frac{219462793027791818759186957}{266727848740241503233353281042665145463487000},\notag\\
	&\frac{14737793385980396424527387159}{277702691081180314127738079406508049818507317200},\frac{76542295754540828585833174473697}{22402592153542599025786896369422989563696466402467600},\notag\\
	&\frac{1394230554274187964861474712183}{6348912756064704177348513141132021333841103838756000},\frac{330978918796758196002883841623103}{23484210185550624295883397889889272366131868197598330800},\dots\Big\}
\end{align*}
}

Finally, we turn to the next order $a^2$ using our proposed correlator at this order. We split into the $\log^2 (x \bar x)$ piece, and the $\log (x \bar x)$ piece. Equating to the SCPW we have
\begin{align}
	F^{(2)}(u,v)|_{\log^2 (x\bar x)} &= \frac {A_l^{(0)}}2 \big(\eta^{(1)}_l\big)^2  G_{2,l} +O(x \bar x)\\
	F^{(2)}(u,v)|_{\log (x\bar x)} &= {\Big(\big(A^{(1)}_{\text{free},l}+A^{(1)}_{\text{int},l}\big)} \eta^{(1)}_l  +  A_l^{(0)} {\eta^{(2)}_l}\Big) G_{2,l} +A_l^{(0)} \big(\eta^{(1)}_l\big)^2  \frac \partial {\partial \gamma} G_{2+\gamma,l} +\mathcal{O}(x \bar x)\ .\label{2nd2}
\end{align}
The first equation indeed yields the correct result  
\begin{align}
	\frac {A_l^{(0)}}2 \big(\eta^{(1)}_l\big)^2=\frac{384 ((l+3)!)^2}{(l+1) (l+6) (2 l+6)!}
\end{align}
as it had to (recall that we derived our result for the string corrected correlator using precisely this consistency condition along with similar for higher twists). Plugging this into the second equation~\eqref{2nd2} and reading off the coefficients of the CPW we obtain from the coefficients of the partial waves, the combination $$\big(A^{(1)}_{\text{free},l}+A^{(1)}_{\text{int},l}\big) \eta^{(1)}_l  +  A_l^{(0)} {\eta^{(2)}_l}\ .$$ Just as for  $A^{(1)}_{\text{int},l}$ itself  we have been unable to obtain a closed formula for this combination, however 
inputting the known coefficients
and rearranging, we obtain values for $\eta^{(2)}_l$ directly, and these are consistent with a simple closed formula (at least for $l>0$)! We find
\begin{align}
	\eta^{(2)}_l = \left\{ \begin{array}{ll}\frac{1344 (l-7) (l+14)}{(l-1) (l+1)^2 (l+6)^2 (l+8)} -\frac{2304 (2 l+7)}{(l+1)^3 (l+6)^3}\qquad &\l=2,4,\dots\\
		\frac{9}{14}\alpha +\frac{1148}{3}\qquad &l=0
	\end{array}\right.\label{2loopanom}
\end{align}
Here $\alpha$ is the remaining undetermined constant in our derivation of  $\langle 2222  \rangle^{(2)}$. We see that surprisingly only the spin zero anomalous dimension depends on the undetermined constant. If we impose the condition $\alpha = 0$ we find that the general formula for $\eta^{(2)}$ holds also for spin zero, however we do not have an independent argument to fix the value of $\alpha$.

\section{Conclusions}

We have bootstrapped  the one-loop amplitude for 4  graviton multiplets in AdS${}_5$ using consistency with the OPE on the dual CFT side. We believe this to be the first such complete one-loop result.
As an ingredient to this computation we computed an entire family of $O(1/N^2)$ anomalous dimensions of double-trace operators of arbitrary spin and twist,  and as an output we computed the $O(1/N^4)$ anomalous dimensions of all twist 4 double trace operators which survive the strong coupling limit.

A number of papers in recent years have discussed general aspects of large $N$ CFTs and their relation to gravitational theories via AdS/CFT from various perspectives. The results here give new concrete data to compare with some of these predictions, and to conclude we will examine a few of these, relating specifically to the twist 4 anomalous dimensions at $O(1/N^4)$.%

 One aspect is the behaviour of anomalous dimensions in the large spin limit.  For example in~\cite{hep-th/0612247,hep-ph/0511302,1502.07707} a non-trivial consistency condition for anomalous dimensions of large spin operators has been shown, following from the reciprocity principle. Following~\cite{1502.07707} in this context, for twist 4 operators,  it can be phrased as follows. Define the Casimir
\begin{align}
	J^2 = (4+ l + a {\eta^{(1)}_{2,l}})(3+ l+a{\eta^{(1)}_{2,l}}) \ ,
\end{align}
where, to the order we are interested in, we need to include the leading order anomalous dimension in the definition of $J^2$, but no further corrections.
Then the claim is that the anomalous dimension has an expansion at large $l$ (equals large $J$) containing only even powers of $1/J$. This was checked to the previous order in~\cite{1502.07707} (where the coupling dependent terms in $J$ could be neglected). Remarkably we find this continues to hold at the next order. Indeed, plugging in the values for the twist 4 anomalous dimensions above, we find that
\begin{align}
	a {\eta^{(1)}_{2,l}}+ a^2 {\eta^{(2)}_{2,l}} = -a\frac{48}{(J^2{-}6) }  +a^2\frac{1344 (J^2{-}110)}{(J^2{-}20) (J^2{-}6)^2}  +O(a^{3})\ .
\end{align}
The key point here is that the anomalous dimension is a rational function of  $J^2$ only without involving the square roots one would expect for an arbitrary function of $l$ expressed in terms of $J$. Note that at leading order, this statement is equivalent to symmetry of the twist 4 anomalous dimensions under $l \rightarrow -l-7$ (since $J^2$ is symmetric under this transformation at leading order). At next order, note that $\eta^{(2)}$ as a function of $l$,  in~\eqref{2loopanom} is written as a sum of two terms.  The first is symmetric under $l \rightarrow -l-7$ and the second antisymmetric. We then see that the antisymmetric term is entirely predicted by $\eta^{(1)}$ together with the above consequence of reciprocity. That our computation indeed agrees with this provides a non-trivial check.

We also observe from~\eqref{2loopanom} that the $O(a^2)$ twist 4 anomalous dimension $\eta^{(2)}_{2,l} =  O(1/l^4)$  at large spin,  whereas at the previous order   $\eta^{(1)}_l =  O(1/l^2)$. Thus in the large $l$ limit, the leading term  receives no $O(a^2)$ corrections. This is consistent with the results of~\cite{Alday:2007mf,1212.3616,1212.4103} that the  coefficient of the leading large $l$ term is related to the $\langle TTT \rangle$ 3-point function which is protected, and its free field value is exactly $O(a)$ with no $a^2$ terms.

Another interesting prediction concerning the twist 4 anomalous dimensions as a function of $l$ is that the function is predicted to be negative, monotonic and convex for spins 2 and higher~\cite{1212.3616,1212.4103,Alday:2017gde}. We find that the twist 4 anomalous dimensions indeed continue to satisfy this property even after including the $O(a^2)$ corrections. Indeed remarkably these hold for all physical values of $a=1/(N^2{-}1)$ that is for all $N\geq 2$ whereas it need only be true for large $N$.
Specifically, for all $N\geq 2$
\begin{align}
	a {\eta^{(1)}_{2,l}}+ a^2 {\eta^{(2)}_{2,l}}&<0 \qquad &\text{for} \qquad l&\geq 2\quad&& \text{(negativity)}\notag\\
	\frac{\partial}{\partial l} \Big(a {\eta^{(1)}_{2,l}}+ a^2 {\eta^{(2)}_{2,l}}\Big)&>0 \qquad &\text{for} \qquad l&\geq 2\quad&& \text{(monotonicity)}\notag\\
	\frac{\partial^2}{\partial l^2} \Big(a {\eta^{(1)}_{2,l}}+ a^2 {\eta^{(2)}_{2,l}}\Big)&<0 \qquad &\text{for} \qquad l&\geq 2\quad&& \text{(convexity)}
\end{align}

As a final comment, in~\cite{Beem:2013qxa,Beem:2016wfs}, numerical bounds on the anomalous dimensions of the twist 4 operators have been found from crossing symmetry in any $\cN=4$ superconformal field theory, as a function of the central charge. For large $N$ our results seem to be consistent with these bounds.%
\footnote{We would like to thank Balt van Rees for comparing with  these bounds.}

\subsection*{Acknowledgements}
JMD and PH  acknowledge a collaboration with Francis Dolan and Hugh Osborn over a decade ago in which we attempted to bootstrap the $O(1/N^4)$ with a similar method to the one used here, but using the known $O(1/N^2)$ anomalous dimensions at twist four only.
FA would like to thank Kostas Skenderis for discussions. 
PH would also like to thank Balt van Rees for many useful and informative discussions.
JMD and HP are supported by the ERC Grant 648630.
PH is supported by an STFC Consolidated Grant ST/L000407/1.

\subsection*{Note Added}

As this paper was being prepared for submission the reference \cite{Alday:2017xua} appeared. We thank the authors of \cite{Alday:2017xua} for discussions on the results on the anomalous dimensions given there at order $a$ and order $a^2$, obtained using a different method. The approach taken there is in perfect agreement with the results on anomalous dimensions quoted here.

\appendix
\section{Protected superconformal blocks}
\label{App-prot}
Following~\cite{1508.03611} we give explicit expressions for the protected superconformal blocks (i.e. half-BPS and semi-short multiplets, see the discussion following equation~(\ref{G2222OPE})). For convenience, let
\begin{align}
	H_\alpha(z):={}_2F_1(\alpha,\alpha,2\alpha,z).
\end{align}
The half-BPS states are then given by 
\begin{align}
F^{\rm half}_{1,0} &=\frac{y\bar{y}}{x\bar{x}}+\frac{\bar{x} (x-y) (x-\bar{y}) H_1(x)-x (\bar{x}-y) (\bar{x}-\bar{y}) H_1(\bar{x})}{x \bar{x} (x-\bar{x})}\notag \\
F^{\rm half}_{2,0}&=\left(\frac{y\bar{y}}{x\bar{x}}\right)^2-\frac{s(x,\bar{x};y,\bar{y})}{(x\bar{x})^2(x-\bar{x})(y-\bar{y})}\times\notag\\
&\qquad\left[\left(\left(\frac{y\bar{y} H_1(\bar{x})+y\bar{x}H_2(\bar{x}) H_{-1}(\bar{y})}{(x-y)}-(y\leftrightarrow\bar{y})\right)-(x\leftrightarrow\bar{x})\right)\right.\notag\\
&\qquad\qquad-\Big(x H_2(x) H_1(\bar{x})-(x\leftrightarrow\bar{x})\Big)\Big(y H_{-1}(\bar{y})-(y\leftrightarrow\bar{y})\Big)\Bigg]
\end{align}
whereas the semi-short multiplets are
\begin{align}
F^{\rm sh.}_{2,l}&=-\frac{s(x,\bar{x};y,\bar{y})}{(x-\bar{x})(y-\bar{y})}\left[\left(\frac{y\bar{y}}{x\bar{x}}\right)^2\left(\left(\frac{\bar{x}^{l+1} H_{-1}(\bar{y}) H_{l+2}(\bar{x})}{\bar{y}^2 (x-y)}-(y\leftrightarrow\bar{y})\right)-(x\leftrightarrow\bar{x})\right)\right. \notag \\
&\qquad\qquad\qquad\qquad\quad-\left.\left(\frac{x^{l-1} H_1(\bar{x}) H_{l+2}(x)}{\bar{x}^2}-(x\leftrightarrow\bar{x})\right)\Big(y H_{-1}(\bar{y})-(y\leftrightarrow\bar{y})\Big) \right]\notag\\
F^{\prime\, \rm sh.}_{2,l}&= \frac{s(x,\bar{x};y,\bar{y})}{(x-\bar{x})(y-\bar{y})}\left[\left(\frac{y\bar{y}}{x\bar{x}}\right)^2\left(\left(\frac{\bar{x}^{l+2} H_{l+3}(\bar{x})}{\bar{y} (x-y)}-(y\leftrightarrow\bar{y})\right)-(x\leftrightarrow\bar{x})\right)\right.\notag\\
&\qquad\qquad\qquad\qquad\quad+\left.\left(\frac{x^l H_2(\bar{x})H_{l+3}(x)}{\bar{x}}-(x\leftrightarrow\bar{x})\right)\Big(y H_{-1}(\bar{y})-(y\leftrightarrow\bar{y})\Big)\right]
\end{align}
and their normalisations read (recall $a=1/(N^2-1)$)
\begin{align}
A_{2,0} &= 8a \notag \\
A_{4,0}&= 2+4a\notag \\
A_{4,l}&= \frac{l! (l+1)! ((l+1) (l+2)+4a)}{(2 l+1)!}\notag \\
A'_{4,l}&=\frac{((l+2)!)^2 ((l+1) (l+4)-12a)}{(2 l+4)!},
\end{align}
for $l$ even and zero otherwise.

\end{document}